%% file: main.tex
\documentclass[conference]{IEEEtran}
\IEEEoverridecommandlockouts

\usepackage{cite}
\usepackage{amsmath,amssymb,amsfonts}
\usepackage{algorithmic}
\usepackage{textcomp}
\usepackage{svg}
\usepackage{amsmath}
\usepackage{xcolor}
\usepackage{graphicx}
\usepackage{graphics}
\usepackage{float}
\usepackage{cclicenses}
\usepackage{xspace}
\usepackage{subfigure}
\usepackage{makecell}
\usepackage{multirow} 

\usepackage[T1]{fontenc}

\usepackage[ruled , linesnumbered]{algorithm2e}
\usepackage{cleveref}

\begin{document}
\title{A Heuristic Approach for Scalable Quantum Repeater Deployment Modeling}
\author{\IEEEauthorblockN{Tasdiqul Islam and Engin Arslan}
\IEEEauthorblockA{University of Nevada Reno\\
tasdiqul@nevada.unr.edu, earslan@unr.edu}}

\maketitle

\newcommand{\name}{$MinRepAlloc$\xspace}
\newcommand{\Lmax}{\mathrm{L}_\mathrm{max}}

\begin{abstract}
\input{abstract}
\end{abstract}

\maketitle

\section{Introduction}
\input{intro.tex}

\section{Background}
\input{background.tex}

\section{Related Work}\label{sec:related}
\input{related_work.tex}

\section{The Heuristic Models}

\input{process.tex}

\section{Evaluations}
\input{result}

\vspace{-2mm}
\section{conclusion and future work}
\vspace{-2mm}
\input{conclusion}

\bibliographystyle{IEEEtran}
\footnotesize

\bibliography{references}
\end{document}

%% file: abstract.tex
Long-distance quantum communication presents a significant challenge as maintaining the fidelity of qubits can be difficult. This issue can be addressed through the use of quantum repeaters to transmit entanglement information through Bell measurements. However, despite its necessity to enable wide-area quantum internet, the deployment cost of quantum repeaters can be prohibitively expensive, thus it is important to develop a quantum repeater deployment model that can strike a balance between cost and effectiveness. In this work, we present novel heuristic models to quickly determine a minimum number of quantum repeaters to deploy in large-scale networks to provide end-to-end connectivity between all end hosts. The results show that, compared to the linear programming approach, the heuristic methods can find near-optimal solutions while reducing the execution time from days to seconds when evaluated against several synthetic and real-world networks such as SURFnet and ESnet. As reliability is key for any network, we also demonstrate that the heuristic method can determine deployment models that can endure up to two link/node failures.


%% file: intro.tex
Quantum networks allow the transmission of quantum bits (aka qubits) between quantum nodes, thus they are crucial for a variety of applications, including quantum key distribution \cite{ekert1991quantumkd,bennett2020quantumkd}, cryptographic communication \cite{pirandola2020advancescrypto}, clock synchronization \cite{jozsa2000quantumclock,chuang2000quantumclock,giovannetti2001quantumclock}, and distributed quantum computing \cite{buhrman2003distributed,beals2013efficientdistributed,cacciapuoti2019quantumdistributed}. However, decoherence presents a significant obstacle for long-distance qubit transmission as it causes the fidelity to deteriorate over time and distance \cite{bergli2009decoherence,saki2019studydecoherence,helm2009quantumdecoherence}. As an example, previous studies showed data a qubit can be transmitted for around $130$ kilometer (around $80$ miles) on a hollow core fiber optic cable before losing its fidelity bellow 90

Entanglement swapping starts with the creation of a pair of entangled qubits on each end of a communication channel. One of the entangled qubits in both endpoints is transferred to a quantum repeater (located between the endpoints), which conducts Bell State Measurement (BSM) using the transferred qubits. The result of BSM is then transferred to one of the endpoints to apply necessary gate operations that will result in the remaining qubits  being entangled. If the distance between endpoints is too far to transmit a qubit from endpoints to a quantum repeater due to decoherence, it is possible to deploy multiple quantum repeaters and repeat the whole process across the intermediate repeaters to create entanglement between any two endpoints regardless of physical distance. As a result, quantum repeaters are essential to enable long-distance quantum communication. 


 In order to establish an initial quantum network, it is crucial to identify a cost-effective approach for deploying the minimum number of quantum repeaters necessary. Determining the optimal placement of quantum repeaters is a complex undertaking, as it requires balancing the performance of the network with the cost of deploying quantum repeaters. As a result, identifying the minimum number of required repeaters and their locations is a critical area of research in the pursuit of scalable quantum networks. In previous work, researchers proposed a linear programming approach to find the quantity as well as the location to deploy quantum repeaters in an existing network topology~\cite{rabbie2022repallocLP}. In this model, system administrators can declare a set of requirements such as the maximum distance that a qubit can be transmitted without losing its fidelity and failure resistance. Then, the model will search for a solution (i.e., the location of quantum repeaters in the network) that can satisfy the requirements using a minimum number of quantum repeaters. We observe that while the linear programming approach is able to find the optimal solution, it is not scalable since its runtime increase exponentially as the scale of the network grows.

In this paper, we propose two heuristic approaches to determine the locations of quantum repeaters in a network that can meet the requirements such as the maximum distance that a qubit can be transmitted and resilience to failures. The first heuristic algorithm, Multi-Center Approach (MCA), starts with identifying the most populated areas of a network to select the location of initial quantum repeaters. It then identifies intermediate nodes to fully connect the initially selected quantum repeater deployment locations. The second heuristic model, Single Center Approach (SCA), chooses one initial location to deploy the first quantum repeater, then determines the next location based on the coverage area of the first one such that it extends the coverage area gradually. 
We find that SCA estimates a similar number of quantum repeaters compared to the resource-intensive Integer Linear Programming (ILP) model~\cite{rabbie2022repallocLP} for different networks. On the other hand, the ILP method takes days to find a solution for a network with $54$ nodes whereas SCA can estimate a solution in less than a second, resulting in several orders of magnitude improvement in execution times.

%% file: background.tex
Similar to traditional networks, the fidelity of a signal (i.e., qubit) decreases in quantum networks as the transfer distance increases~\cite{chakraborty2019distributed}. Traditional networks rely on repeaters that measure the signal to amplify it for the next segment of a link. However, \emph{no cloning} theorem prevents the creation of a perfect copy of a qubit, hence the traditional approach of measure-then-reproduce cannot be applied to strengthen the qubits during their transmission. Quantum repeaters thus employ a different approach that utilizes Bell state measurement to transmit the entanglement between endpoints. 

\begin{figure}
\begin{center}
\includegraphics[keepaspectratio=true,angle=0,width=0.8\linewidth] {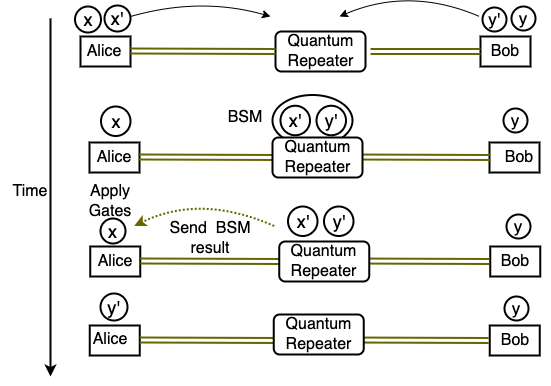}
\vspace{-3mm}
\caption{Illustration of entanglement swapping using a quantum repeater.}
\label{fig:entanglement_swapping}
\vspace{-6mm}
\end{center}
\end{figure} 
Figure~\ref{fig:entanglement_swapping} illustrate this process. If Alice and Bob are located far from the maximum distance a qubit can be transmitted without losing its fidelity, they create entangled qubit pairs as $x,x'$ and $y,y'$. They then send one of the entangled qubits, $x'$ and $y'$, to the quantum repeater that is located between them. The repeater executes Bell State Measurement on $x'$ and $y'$ and sends the result to Alice using a classical communication channel. Finally, Alice applies gate operations to $x$ based on BSM results which causes qubits $x$ and $y$ to be entangled. If the distance between Alice and Bob is longer than a single repeater can handle, the same process (transfer a qubit between adjacent nodes, perform BSM, and share the results via classical channel) is repeated over multiple intermediate nodes to extend the range of entanglement and enable long-distance quantum communication. 


%% file: related_work.tex
\cite{yao2021optimaldeploymentdesign} showed that adding quantum repeaters without proper planning can lead to inefficient resource utilization and reduced data transfer rates. For long-distance quantum entanglement distribution, \cite{liorni2021quantumrepinspace} proposed a method based on the integration of satellite-based optical links and quantum repeaters. Finally, there are numerous works on routing algorithms and protocols \cite{chakraborty2019distributed,dahlberg2019linklayerprot,kozlowski2020designingqnprot,yang2022online,zhao2021redundant,zhao2022segmented} that consider established quantum networks with the necessary number of quantum repeaters.

Rabbie et al. developed a quantum repeater deployment model on top of existing network infrastructure~\cite{rabbie2022repallocLP}. Instead of 
 building completely a new infrastructure from scratch, the proposed solution identifies existing network hubs that can be used to deploy quantum repeaters. Thus, it takes existing network infrastructure as input and identifies a set of vertices (i.e., the location of traditional network routers) to colocate quantum repeaters such that end-to-end quantum connectivity can be achieved using existing  network connectivity (e.g., fiber optical cables) and deployed quantum repeaters. The authors utilize  Integer Linear Programming (ILP) to find the optimal deployment locations. However, the computational cost of the developed ILP method increases exponentially with increasing network scale. Thus, the authors only tested the proposed method using small-scale network with up to $30$ vertices.

Quantum repeater deployment is similar to the efficient placement of electric vehicle charging stations. In \cite{lam2014electriccomplexity}, the complexity of various methods for placing charging stations is discussed. The authors propose two Integer Linear Program approaches, one greedy approach, and one chemical reaction optimization method. They demonstrate that the ILP method can determine a better optimal number of charging stations, but it cannot converge to a solution when the number of nodes is more than $200$ nodes, severely restricting its applicability for today's large-scale networks. This work inspires us to develop heuristic models to determine the near-optimal number of quantum repeaters. 

%% file: process.tex
\begin{figure}[t]
\begin{center}
\includegraphics[keepaspectratio=true,angle=0,width=0.6\linewidth] {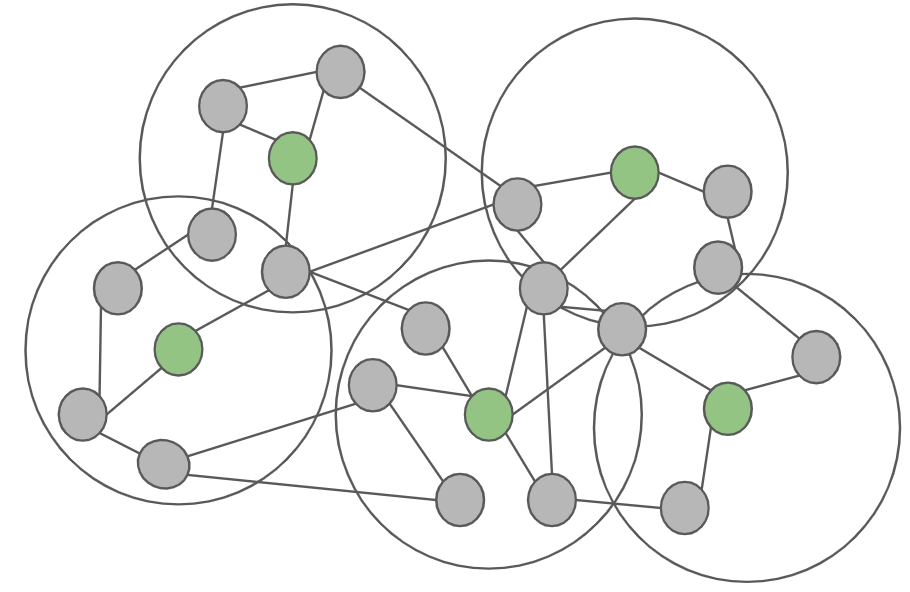}
\vspace{-3mm}
\caption{Choosing initial nodes (i.e., centers) for quantum repeater deployment modeling based on Multi-Center Approach.}
\label{fig:choosing_centers1}
\vspace{-6mm}
\end{center}
\end{figure} 
As we aim to utilize existing network hubs and links, we take existing network topology as a graph that consists of vertices and edges and find a set of vertices to deploy quantum repeaters. Obviously, the brute force approach is intractable as its execution time increases exponentially. We show in evaluations that linear programming also suffers from extremely long execution times. Thus, we introduce two novel heuristic solutions to tackle this problem efficiently in terms of computational time and the number of quantum repeaters to e deployed. We first define \emph{coverage area} for a quantum repeater as the area in which it can provide quantum entanglement swapping, $L_{max}$. Figure~\ref{fig:choosing_centers1} illustrates this concept for the vertices colored in green. The circles represent the area of green vertices that can serve directly without additional quantum repeaters. All the vertices within the coverage area must be less than $L_{max}$ distance away from the center vertex. Hence the coverage area is not guaranteed to be a complete circle since we consider the graph path when calculating the distance between the center to other nodes.   

The first approach, \emph{Multi Center Approach (MCA)}, identifies a minimum set of vertices whose coverage area will encapsulate all other vertices in the network, as shown in Figure~\ref{fig:choosing_centers1}. These vertices are added to the list of selected vertices for quantum deployment, $V_{QR}$. Since the graph path distance between the vertices in $V_{QR}$ can be longer than $L_{max}$, it then finds checks accessibility between the center nodes and adds them to $V_{QR}$ to provide complete end-to-end coverage for all vertices in the network. The second approach, aka \emph{Single Center Approach (SCA)}, on the other hand, adopts a gradual coverage expansion scheme. It identifies the first vertex with the most extensive coverage area and adds it to the $V_{QR}$ list. It then selects a new vertex at the periphery of the coverage area of the first vertex to add it $V_{QR}$, thereby gradually expanding the coverage area of the vertices in the $V_{QR}$ list. 

\newcommand{\Vleaf}{\mathrm{V}_\mathrm{leaf}}
\newcommand{\Cmandatory}{\mathrm{V}_\mathrm{access}}
\newcommand{\Vremaining}{\mathrm{V}_\mathrm{remaining}}
\newcommand{\Vcovered}{\mathrm{V}_\mathrm{covered}}
\setlength{\textfloatsep}{1pt}
\begin{algorithm}[t]
\caption{Algorithm to choose initial vertices to deploy quantum repeaters (MCA)}\label{alg:mca_center_selection}
\KwData{$G = (V , E ,L)$}
\KwResult{C = Set of center nodes}
\SetKwFunction{ChooseCenters}{ChooseCenters}
\SetKwProg{Fn}{Function}{:}{}
\Fn{\ChooseCenters{$G$}}{
    $\Vleaf \gets \{n \in V , deg(n)=1\}$\;
    $\Cmandatory \gets \{n \in V , E(n , l) = true , l \in \Vleaf \}$\;
    \For{ $v \in \Cmandatory$}{
                $v_{leaf} \gets \{n \in \Vleaf , E(n , v) = true\}$\;
                \If{$\exists L(\mathrm{v_1} \in v_{leaf} , \mathrm{v_2} \in v_{leaf} ) > \Lmax$}{
                    $C \gets C \cup v$\;
                    \For{ $v' \in coverage(v)$}{
                        $\mathrm{V}_\mathrm{covered} \gets \mathrm{V}_\mathrm{covered} \cup v'$ \;
                   }
                }
        }
    $\Vremaining \gets V \setminus \mathrm{V}_\mathrm{covered} $\;
    \While{$\Vremaining \ne \varnothing$}{
        $v \gets max ( coverage(v), v \in V_{remaining})$\;
        $C \gets C \cup v$\;
        \For{ $v' \in coverage(v)$}{
            $\mathrm{V}_\mathrm{covered} \gets \mathrm{V}_\mathrm{covered} \cup v'$ \;
        }
        $\Vremaining \gets V \setminus \mathrm{V}_\mathrm{covered} $\;
       }
    \textbf{return} C;
}
  \textbf{End Function}
\end{algorithm}

\subsection{Multi Center Approach (MCA)}
MCA consists of two steps as (i) center selection and (ii) center connection. The center selection step involves identifying a set of vertices whose combined coverage area encapsulates all the vertices in the network. Algorithm~\ref{alg:mca_center_selection} provides the pseudo-code for the center selection. We first label nodes\footnote{Words node and vertex used interchangeably in the rest of the paper.} that are connected to only one other node as leaf nodes, $V_{leaf}$. We next define access nodes, $V_{access}$, connected to leaf nodes. Access nodes are important since they provide connectivity to leaf nodes; hence, we add access nodes to the list of center nodes, $C$, if the distance between their leaf nodes is larger than $\Lmax$. In other words, if two leaf nodes sharing the same access node want to communicate with each other, then the access node must be selected for a quantum repeater unless the distance between them (calculated by $L(l1,l2)$) is shorter $\Lmax$. After adding the eligible access nodes to the list of center nodes, we check if all the vertices are covered by the coverage area of at least one center vertex. If not, we add additional vertices to the center list. In this step, we only consider nodes not currently covered by any center nodes. We calculate the coverage score for each eligible node which refers to the number of uncovered nodes that fall within the $L_{max}$ distance of a node. Once the coverage score is calculated, we pick the node with the highest score to add it to the center node list and mark all other nodes within $L_{max}$ distance as covered. We repeat this process until all nodes are covered.



After selecting the center nodes, the next step involves checking for connectivity between center nodes since it is possible that the length of the route between center nodes can exceed the maximum entanglement distance, $L_{max}$. We propose two solutions to select intermediate nodes to connect the center nodes selected in the first step. The first approach calculates Minimum Spanning Tree (MST) (using Kruskal’s Algorithm~\cite{kruskal1956shortest}) using the center nodes, $C$, then iterates over all the links found by MST. If a link is longer than $\Lmax$, then we add nodes along the link to the intermediate node list, $I$, such that they will be used for quantum repeaters. Please note that this approach only considers the nodes located on the links of MST (i.e., the shortest path between the center node); thus, it does not consider all possible nodes. The second approach intends to overcome this limitation by considering all possible commonly accessible nodes between center nodes. Specifically, it again calculates MST but then checks all nodes that are accessible by the neighboring center nodes such that it can potentially reuse the same intermediate node to achieve connectivity between multiple center nodes. 

\begin{algorithm}[t]
\caption{Algorithm to find intermediate nodes on graph path to connect the center nodes (MCA-GP).}
\label{alg:mca_graph_path}
\KwData{$G = (V , E ,L)$}
\KwResult{I = Set of intermediate nodes}
$C \gets \ChooseCenters{G}$\;
$MST \gets MinimumSpanningTree(C)$\;
$I \gets \varnothing$\;
\For{edge in $MST$}{
$nodes = getListofNodes(edge)$;\
$node1 = \gets nodes[0]$\;
\For{ $i=1$ \KwTo length(nodes) - 1}{
            $node2 \gets nodes[i]$\;
            \eIf{$L(node1 , node2) > \Lmax$}{
                $I \gets I \cup nodes[i -1]$\;
                $node1 \gets nodes[i-1]$\;
            }{
            
            }
    }
}
return I;
\end{algorithm}

\begin{figure}[t]
\begin{center}
\includegraphics[keepaspectratio=true,angle=0,width=0.6\linewidth] {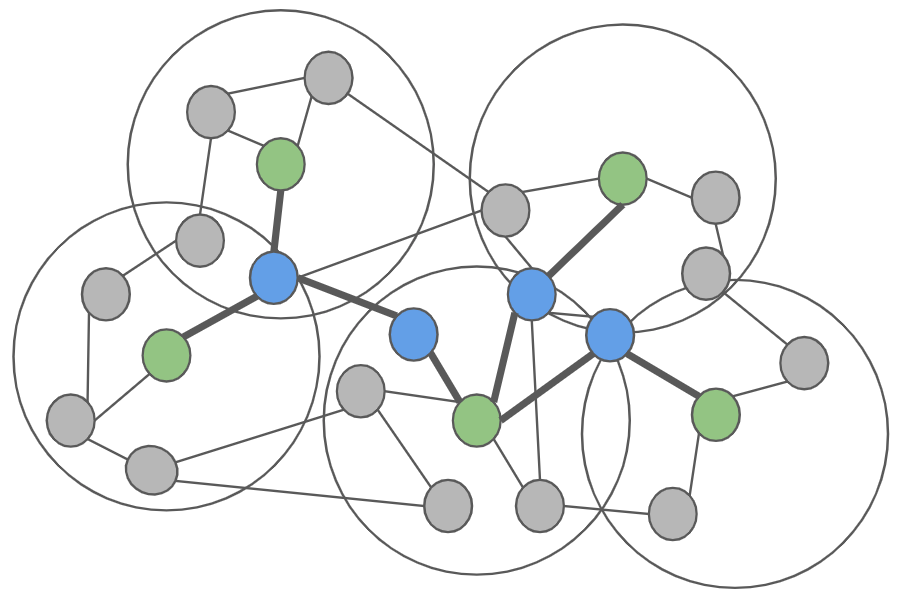}
\vspace{-3mm}
\caption{Illustration of intermediate node selection process to provide connectivity between center nodes.}
\label{fig:intermediate_nodes}
\vspace{-6mm}
\end{center}
\end{figure} 
\textit{Graph Path-Based Intermediate Node Selection (MCA-GP):} In this method, we first calculate the Minimum Spanning Tree (MST) connecting the center nodes, then check if there exist repeaters along the MST path between the two center nodes as shown in Algorithm~\ref{alg:mca_graph_path}. If there are no available repeaters or the distance exceeds the $\Lmax$, we select additional nodes along the path for quantum repeater deployment to establish a feasible path between the center nodes as illustrated in Figure~\ref{fig:intermediate_nodes}. If the distance between centers exceeds $\Lmax$ and there are no other nodes in between, we define new nodes and add them to the list since the nodes; otherwise, hosts cannot be connected using the existing links. Finally, we verify the correctness of the solution by finding a path between each pair of nodes in the network that is completely covered by quantum repeaters. Therefore, all nodes in the topology can establish quantum communication with all other nodes.

\begin{algorithm}[tb]
\caption{Algorithm to select intermediate nodes considering nodes on non-graph path routes (MCA-Flex)}\label{alg:mca_ngp}
\KwData{$G = (V , E ,L)$}
\KwResult{I = Set of intermediate nodes}
$C \gets \ChooseCenters{G}$\;
$H \gets subgraph(G) $ , with nodes only C\;
$T \gets MST(H)$\; \label{line:mca_ngp_mst}
$\mathrm{E}_\mathrm{mst} \gets \mathrm{E}_\mathrm{T}$ \;
$I \gets \varnothing$ \;
$common\_node\_map \gets \varnothing$ \;
\For{$i,j \in \mathrm{E}_\mathrm{mst}$}{
$common\_nodes \gets \{u: L(i , u) \leq \Lmax\} \cap \{v: L(j , v) \leq \Lmax\}$ \; \label{line_mca_ngp_common}
\For{$n \in common\_nodes$}{ \label{line_mca_ngp_cmn_node_map_start}
$common\_node\_map[n] \gets common\_node\_map[n].add(pair(i , j))$ \;

} \label{line_mca_ngp_cmn_node_map_end}
}
$sort(common\_node\_map)$ \; \label{line_mca_ngp_sort_map}
\For{$n \in common\_node\_map$}{ \label{line_mca_ngp_select_node_start}
    $pair\_list \gets common\_node\_map[n]$ \;\label{line:test}
    \For{$pair \in pair\_list$}{
        $\mathrm{E}_\mathrm{mst}.remove(pair)$ \;
       
    }
    $I \gets I \cup n$\; \label{line_mca_ngp_pick_int_node}
     \If{$\mathrm{E}_\mathrm{mst} == \varnothing$}{
        break;
        }
} \label{line_mca_ngp_select_node_end}
return I;

\end{algorithm}

\begin{figure}[t]
\begin{center}
\includegraphics[keepaspectratio=true,angle=0,width=0.8\linewidth] {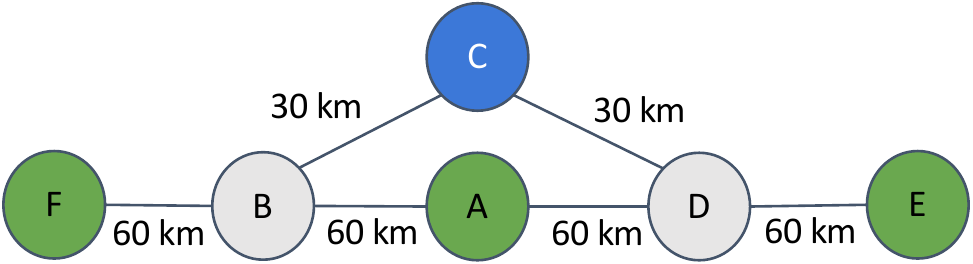}
\vspace{-2mm}
\caption{Considering intermediate nodes outside the graph path can reduce the number of quantum repeaters needed. While two quantum repeaters, located on B and C, are needed to connect node A to nodes E and F when restricting the search to the graph path, only one node, C, will be sufficient to provide the same connectivity. $\Lmax=100$ km.}
\label{fig:mca-flex-example}
\vspace{-2mm}
\end{center}
\end{figure} 

\textit{Flexible Intermediate Node Selection (MCA-Flex):} 
Since the MCA-GP only considers nodes along the graph path (i.e., cycles are not allowed) when connecting the center nodes, it ignores other nodes that multiple center nodes can access. Figure~\ref{fig:mca-flex-example} demonstrates this in a network where nodes $A$, $F$, and $E$ are selected as center nodes with Algorithm~\ref{alg:mca_center_selection} with $\Lmax=100km$. The  MCA-GP would choose node $B$ to connect centers $A$ and $F$ and node $D$ to connect $A$ and $E$. However, choosing only node $C$ as a quantum repeater is sufficient to establish a quantum network route between nodes $A$ and $E$ and $A$ and $F$. However, since both A-C-E and A-C-F involve loops (e.g., the route between A and E will travel A-D-C-D-E), they are not considered in the  MCA-GP.

In Algorithm-\ref{alg:mca_ngp}, we take the MST of the centers in line \ref{line:mca_ngp_mst}. Then, in line \ref{line_mca_ngp_common}, we select the common nodes of every pair of nodes of the MST edges. In line \ref{line_mca_ngp_cmn_node_map_start} to \ref{line_mca_ngp_cmn_node_map_end}, we maintain a map(common\_node\_map) that records the list of center pairs that share the same node. In line \ref{line_mca_ngp_sort_map}, we sort the common\_node\_map based on the number of center pairs sharing a node. We pull the node with the largest number of shared center pairs from the common\_node\_map list and remove it from the list in line \ref{line_mca_ngp_select_node_start} to \ref{line_mca_ngp_select_node_end}. This is to ensure that if there is a common node between a pair of center nodes, then these center nodes are guaranteed to communicate with each other. We repeat this process until all pairs are covered.

\begin{algorithm}[t]
\caption{Single Center Approach to find nodes for Quantum Repeater Deployment (SCA)}\label{alg:single_center_approach}
\KwData{$G = (V , E ,L)$}
\KwResult{C = Set of center nodes}
\SetKwFunction{ChooseCenters}{ChooseCenters}
\SetKwProg{Fn}{Function}{:}{}
\Fn{\ChooseCenters{$G$}}{
$C \gets C \cup n, max ( \mathrm{n}_\mathrm{in\_circle} )  $\; \label{line_sca_first_node}
$\Vcovered \gets {v: L(v ,c) \leq \Lmax, v \in V, c \in C}$ \;
\label{line_sca_covered_nodes}
$\Vremaining \gets V \setminus \mathrm{V}_\mathrm{covered} $

\While{$\Vremaining \ne \varnothing$}{ \label{line_sca_next_node_start}
    $max\_covered \gets 0$ \;
    $next\_center \gets \varnothing$ \;
    \For{$n \in \Vcovered$}{
        \If{$length(\{v: L(v ,n) \leq \Lmax\}) \geq max\_covered$}{ \label{line_sca_check_high_score_start}
        $max\_covered = length(\{v: L(v ,n) \leq \Lmax\})$ \;
        $next\_center \gets n$ \;
        } \label{line_sca_check_high_score_end}
    }
    $C \gets C \cup next\_center  $\;
    $\Vremaining \gets \Vremaining \setminus \{v: L(v , next\_center) \leq \Lmax , v \in \Vremaining \}$  
  } \label{line_sca_next_node_end}
  \textbf{return} C;
  }
   \textbf{End Function}
  
\end{algorithm}

\begin{figure*}
\begin{center}
\subfigure[Step 1]{
\includegraphics[keepaspectratio=true,angle=0,width=.13\linewidth] {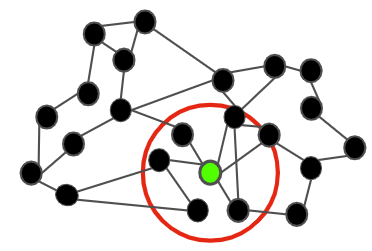}
\label{fig:center2_1}}
\hspace{-4mm}
\subfigure[Step 2]{
\includegraphics[keepaspectratio=true,angle=0,width=.13\linewidth] {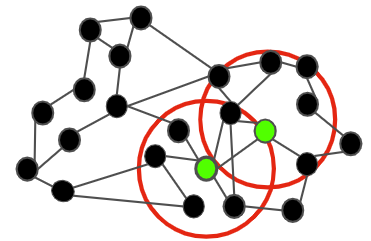}
\label{fig:center2_2}}
\hspace{-4mm}
\subfigure[Step 3]{
\includegraphics[keepaspectratio=true,angle=0,width=.13\linewidth] {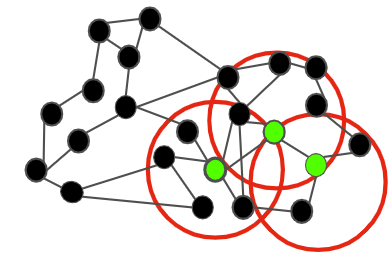}
\label{fig:center2_3}}
\hspace{-4mm}
\subfigure[Step 4]{
\includegraphics[keepaspectratio=true,angle=0,width=.13\linewidth] {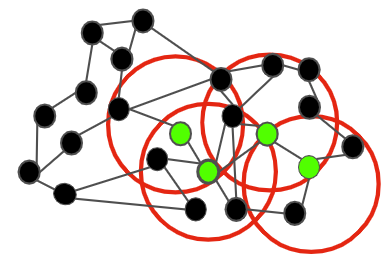}
\label{fig:center2_4}}
\hspace{-4mm}
\subfigure[Step 4]{
\includegraphics[keepaspectratio=true,angle=0,width=.13\linewidth] {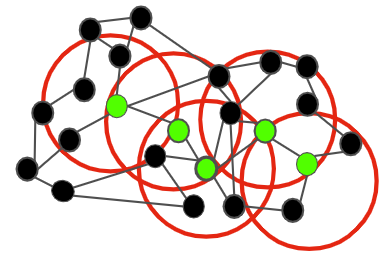}
\label{fig:center2_5}}
\hspace{-4mm}
\subfigure[Step 6]{
\includegraphics[keepaspectratio=true,angle=0,width=.13\linewidth] {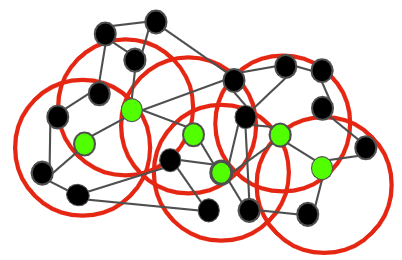}
\label{fig:center2_6}}
\hspace{-4mm}
\subfigure[Step 7]{
\includegraphics[keepaspectratio=true,angle=0,width=.13\linewidth] {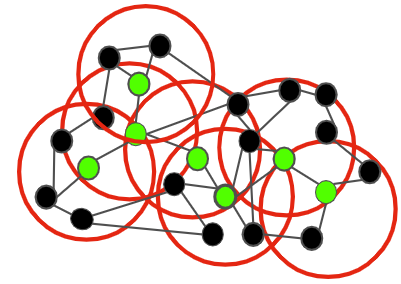}
\label{fig:center2_7}}

\caption{Illustration of how Single Center Approach (SCA) selects nodes to deploy quantum repeaters. It starts with one node and expands the coverage by choosing another node that is near the edge of the current coverage area.}
\vspace{-6mm}
\label{fig:center2}
\end{center}
\end{figure*} 
\subsection{Single Center Approach (SCA)}
MCA follows a two-step approach to identify the center nodes and intermediate nodes. Thus, it may result in a suboptimal solution if these steps are not planned together. In other words, if the center node selection policy does not consider the connectivity needs between the center nodes, it may need to choose more intermediate nodes to provide connectivity between the center nodes.  Instead, if we choose centers in a way that there is already a feasible distance between them, it would further reduce the total number of required quantum repeaters. Thus, we introduce Single Center Approach (SCA) that picks only one center node and increases the coverage area by selecting new nodes near the first selected center node as detailed in Algorithm \ref{alg:single_center_approach}. As illustrated in Figure \ref{fig:center2}, we choose the next center node that is not more than $L_{max}$ away from the previously chosen center node. This way, we do not have to choose intermediate nodes as quantum repeaters. We select the node that covers the maximum number of uncovered nodes and repeat this process until all the nodes are covered.

In line \ref{line_sca_first_node} of Algorithm~\ref{alg:single_center_approach},  we choose a node to be the first quantum repeater whose coverage area encapsulates the maximum number of nodes. In line \ref{line_sca_covered_nodes}, we mark the nodes close to the center node with distance $\Lmax$ as covered nodes and start expanding the coverage area. In line \ref{line_sca_next_node_start} to \ref{line_sca_next_node_end}, we choose the next node based on coverage score (i.e., the number of uncovered nodes that are within distance $\Lmax$ of a given node) of all eligible nodes. We then take the node with the highest coverage score and repeat the process until all nodes are covered.

\begin{algorithm}[t]
\caption{Algorithm to choose center nodes with robustness factor $K$}\label{alg:mca_center_selection_robust}
\KwData{$G = (V , E ,L)$ , K}
\KwResult{C = Set of center nodes}
    $C \gets \varnothing$\;
    \For{$i=1$ \KwTo $K$}{
    $\mathrm{C}_\mathrm{i} \gets \ChooseCenters{G}\}$ \;
    $C \gets C \cup \mathrm{C}_\mathrm{i}$ \;
    $G.remove\_nodes(\{v: v \in \mathrm{C}_\mathrm{i}\})$ \;
    \For{$v \in V$}{
        \If{$length(\{c: L(c , v) \leq \Lmax, c \in C\}) \geq K$}{
            $G.remove\_node(v)$ \;
        }
    }
}
return C;
\end{algorithm}
\subsection{Failure Resistance}
As node and link failures occur in the networks, building quantum networks with reliability in mind is essential. We, thus, look into the availability of backup routes for each connection to provide resilience against node and link failures. Rabbie et al. ~\cite{rabbie2022repallocLP} defined the "robustness parameter" to refer to the number of distinct routes between any pair of nodes.  We adopt the same metric and extend both MCA and SCA algorithms to find the nodes to deploy quantum repeaters that can satisfy the robustness requirement. 

\begin{algorithm}
\caption{Algorithm to choose intermediate nodes with MCA-Flex considering robustness factor K)}\label{alg:mca_ngp_robust}
\KwData{$G = (V , E ,L)$, C = Set of center nodes, k = Robustness parameter}
\KwResult{I = Set of intermediate nodes}
$H \gets subgraph(G) $ , with nodes only C\;
$T \gets mst(H)$\;
$\mathrm{E}_\mathrm{mst} \gets \mathrm{E}_\mathrm{T}$ \;
$I \gets \varnothing$ \;
$common\_node\_map \gets \varnothing$ \;
\For{$i,j \in \mathrm{E}_\mathrm{mst}$}{
$common\_nodes \gets \{u: L(i , u) \leq \Lmax\} \cap \{v: L(j , v) \leq \Lmax\}$ \;
\For{$n \in common\_nodes$}{
$common\_node\_map[n] \gets common\_node\_map[n].add(pair(i , j))$ \;
}
$edge\_map[pair(i,j)] \gets k$ \;
}
$sort(common\_node\_map)$ \;

\For{$n \in common\_node\_map$}{ \label{line_sca_k_pick_node_start}
    $pair\_list \gets common\_node\_map[n]$ \;
    \For{$pair(i,j) \in pair\_list$}{
        $edge\_map[pair(i,j)] \gets edge\_map[pair(i,j)] -1 $ \; \label{line_mca_ngp_robust_decease_k}
    }
    $I \gets I \cup n$\;
    \If{$edge\_map[pair(i,j)] \leq 0$}{
         delete(edge\_map[pair(i,j)])\;
    }
    \If{$length(edge\_map) == 0$}{
        break\;
    }
    
} \label{line_sca_pick_node_end}
return I;
\end{algorithm}
For MCA, we first modify the center selection algorithm to repeat the center node selection process $K-1$ times, excluding the previously chosen center nodes as shown in Algorithm \ref{alg:mca_center_selection_robust} . This ensures that  all nodes are covered by at least $K$ quantum repeaters such that even if one of them fails, the nodes can route their traffic using the remaining repeaters. We next extend the flexible intermediate node selection algorithm (MCA-Flex) (i.e., Algorithm~\ref{alg:mca_ngp}) to find multiple alternative routes between the center nodes to offer resilience against possible failures as shown in Algorithm \ref{alg:mca_ngp_robust}. The modified algorithm first obtains the edge list from the minimum spanning tree of the center nodes. Then it finds $K$ intermediate nodes between every edge where the distance between the edge nodes is more than $\Lmax$.

Specifically, it identifies common nodes for pairs of every center node pair for which the distance is greater than $\Lmax$. It then creates a key-value object for each node where the node is the key and a list of center nodes for which the node is within the coverage area as a value. In other words, the value holds the list of center pairs for which the node falls within the coverage area of both center nodes. It then sorts the map of nodes based on the number of pairs they are common in descending order. Next, it creates another map of the pairs from the MST edges and assigns a value of k to each one. It selects a node from the sorted map and decrements the value associated with its parent pair in the later map. If the associated value becomes 0, we remove that parent from the map.  Finally, if any pair of center nodes is left with less than $K$ common nodes, we process that pair according to Algorithm \ref{alg:mca_graph_path}. We select the in-between nodes of the remaining pairs in the map, along with $K$ redundant nodes, as quantum repeaters. To extend SCA with robustness, we execute Algorithm~\ref{alg:single_center_approach} $K$ times, each time ignoring nodes selected in previous rounds. Doing so returns $K$ different quantum repeater solutions for each node, ensuring reliability against $K-1$ failures. 

\subsection{Time Complexity Analysis}
The complexity shortest path calculation between all pairs of nodes is $\mathcal{O}(V^2)$ with the Dijkstra Algorithm where $V$ is the number of nodes. The center selection method for MCA (Algorithm~\ref{alg:mca_center_selection}) has the time complexity of $\mathcal{O}(V + E$) for finding leaf nodes where $E$ is the number of edges as it involves finding the degree of each vertex. Then, finding center nodes from access nodes has $\mathcal{O}(V^2)$ complexity. Finding the coverage area of all nodes has $\mathcal{O}(V^2)$ time complexity. So, the total time complexity is $\mathcal{O}(V^2 + E))$.

The code to select intermediate nodes to connect center nodes (i.e., Algorithm \ref{alg:mca_graph_path}) has a time complexity of $\mathcal{O}(E\log{}V)$ for minimum spanning tree calculation using the Kruskal's Algorithm. Then, finding intermediate nodes along all pairs of centers takes $\mathcal{O}(VE)$, which brings the time complexity of Algorithm \ref{alg:mca_ngp} to $\mathcal{O}(V(V+E))$. Consequently, the time complexity of MCA-GP is $\mathcal{O}(V(V+E))$. The time complexity reduces to $\mathcal{O}(V^2 + E\log{}V)$ when considering links outside of the graph path (i.e., MCA-Flex).


We also calculate the shortest path for SCA (Algorithm \ref{alg:single_center_approach}); thus, it also takes $\mathcal{O}(V^2)$ for that operation. To find nodes with maximum coverage score, we calculate the coverage score for every node in every iteration, which leads to $\mathcal{O}(V^2)$ time complexity. So, the complexity of SCA becomes $\mathcal{O}(V^2)$. Since the robustness parameter $K << V$, the time complexity for the algorithms that consider the robustness factor is the same as the original algorithms for the complexity remains the same. As a result, compared to ILP, whose runtime increases exponentially as input data grows, the heuristic methods offer a much faster option with polynomial time complexity.  

%% file: result.tex
\begin{figure}
\begin{center}
\includegraphics[keepaspectratio=true,angle=0,width=.95\linewidth] {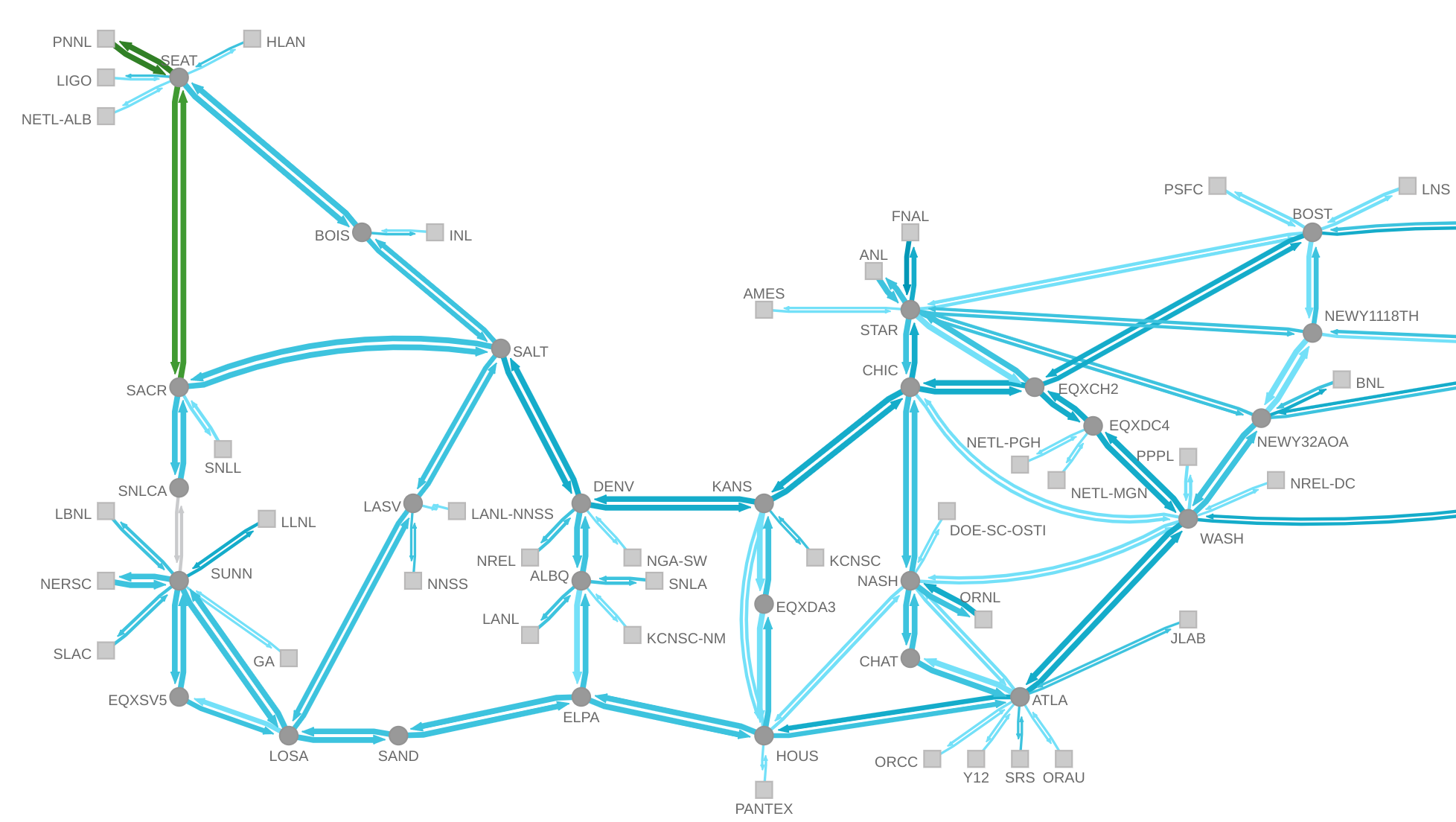}
\vspace{-3mm}
\caption{ESnet consists of $63$ nodes. Since some links are longer than $\Lmax$, we augmented it with additional nodes to find a feasible quantum repeater deployment solution that covers the entire network.}
\label{fig:ESnet}
\vspace{-4mm}
\end{center}
\end{figure}

\begin{table}
\renewcommand{\arraystretch}{1.3}
\centering
\caption{Comparison of heuristic approaches against ILP~\cite{rabbie2022repallocLP} method for SURFnet topology in terms quantum repeater count.}
\vspace{-2mm}
\label{tab:proposed_models_compare}
\begingroup
\setlength{\tabcolsep}{2pt} 
\begin{tabular}{c c c c c}
\hline
Maximum transfer   &  \multicolumn{4}{c}{Quantum Repeaters Needed}\\
\cline{2-5}
 distance ($\Lmax$)  & MCA-GP & MCA-Flex & SCA & ILP\\
\hline
130 km  & 4  & 4 & 3 & 3\\
\hline
100 km   & 8 & 7 & 5 & 4\\
\hline
80 km   & 13 & 12 & 8 & 7\\
\hline
60 km   & 22 & 14 & 13 & 12\\
\hline
50 km  & 26 & 19 & 19 & 18 \\
\hline
40 km  & 34 & 31 & 27 & 26\\
\hline
\end{tabular}
\endgroup
\end{table}

We test the proposed solutions using one small synthetic network and two real-world networks; SURFnet~\cite{surfnet} and ESnet~\cite{esnet}. The synthetic network contains only  $14$ nodes. The SURFnet contains $54$ nodes with a distance between the nodes is typically less than $30$km. On the other hand, the ESnet Layer-3 topology has $63$ nodes excluding the one located outside the US (illustrated in Figure~\ref{fig:ESnet}), with the distance between some nodes reaching $898$ km . We ran the simulations using a server with a $128$ core AMD EPYC 2.6 GHz CPU and $1$ TiB main memory. Since all algorithms are not designed to take advantage of multiple cores, they only utilize one core during the execution. To simulate users connected to any node in the cluster, we added a ``ghost'' node to every node in the network with negligible distance. We compare the heuristic models against the Integer Linear Programming (ILP) approach proposed in~\cite{rabbie2022repallocLP}. Unlike ILP which is designed to find solutions for a given set of end nodes, both MCA and SCA return solutions that would allow all the endpoints (i.e., ghost nodes) to communicate.

\begin{figure*}
\begin{center}
\subfigure[ILP-$\Lmax=130km$]{
\includegraphics[keepaspectratio=true,angle=0,width=.23\linewidth] {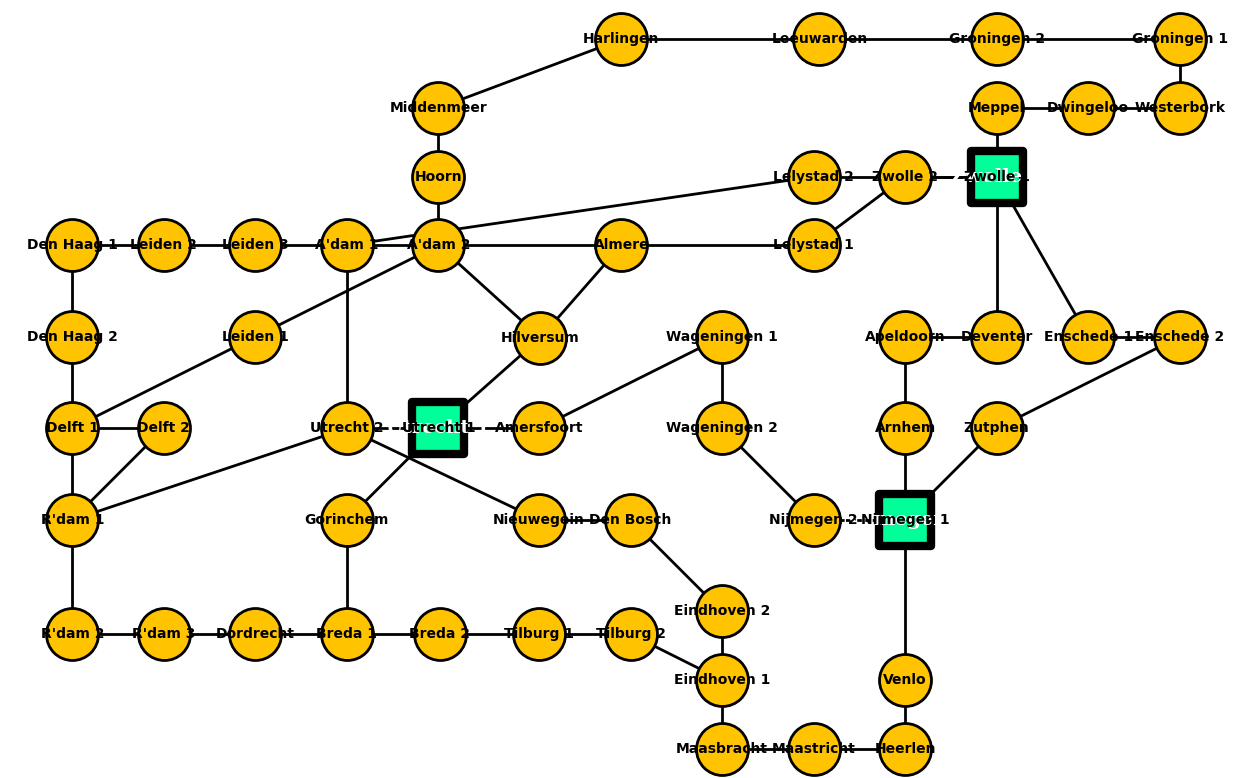}
\label{fig:lp130Surfnet}}
\hspace{-4mm}
\subfigure[ILP-$\Lmax=100km$]{
\includegraphics[keepaspectratio=true,angle=0,width=.23\linewidth] {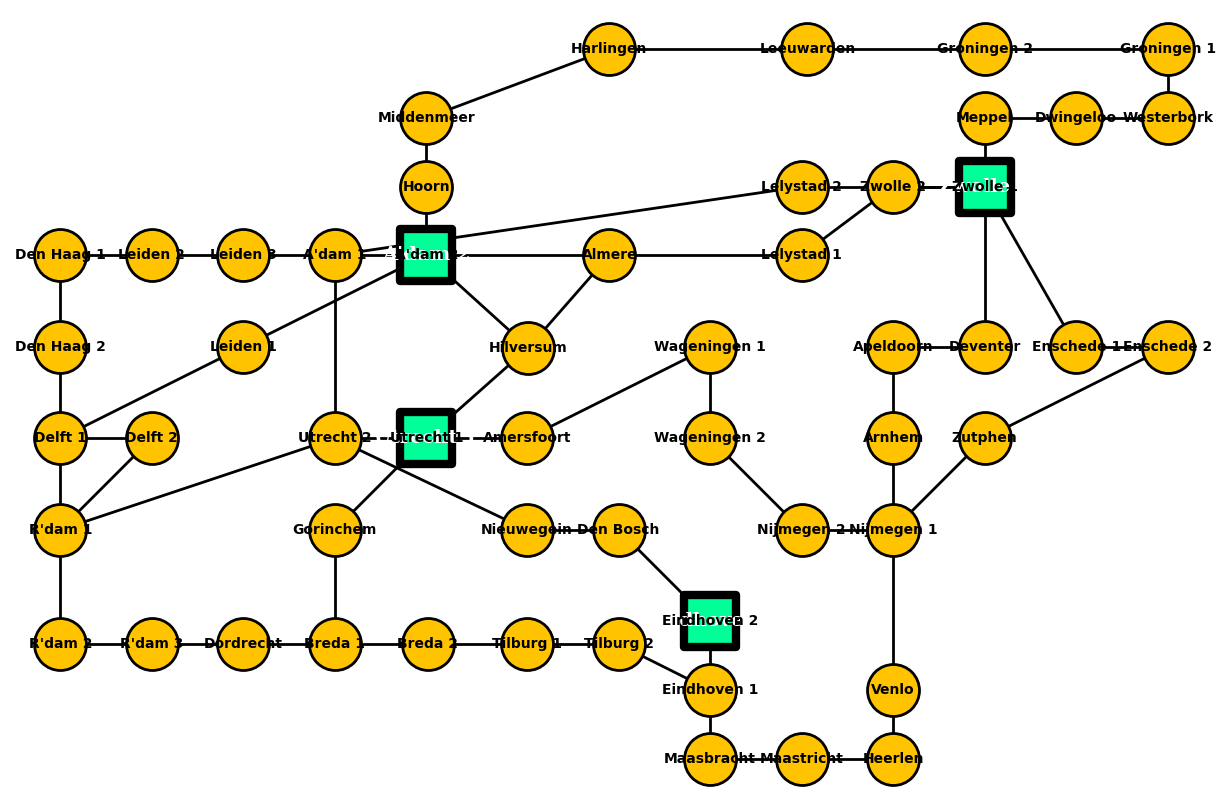}
\label{fig:lp100Surfnet}}
\hspace{-4mm}
\subfigure[ILP-$\Lmax=80km$]{
\includegraphics[keepaspectratio=true,angle=0,width=.23\linewidth] {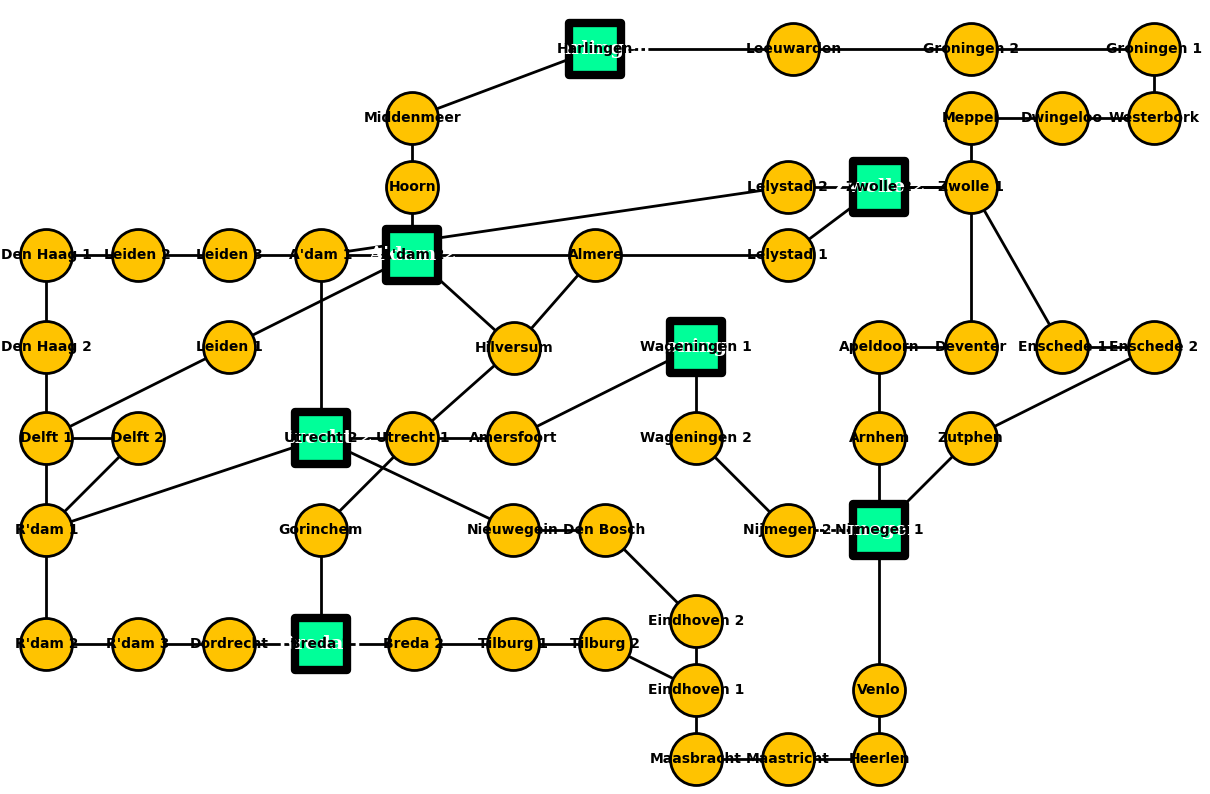}
\label{fig:lp80Surfnet}}
\hspace{-4mm}
\subfigure[ILP-$\Lmax=60km$]{
\includegraphics[keepaspectratio=true,angle=0,width=.23\linewidth] {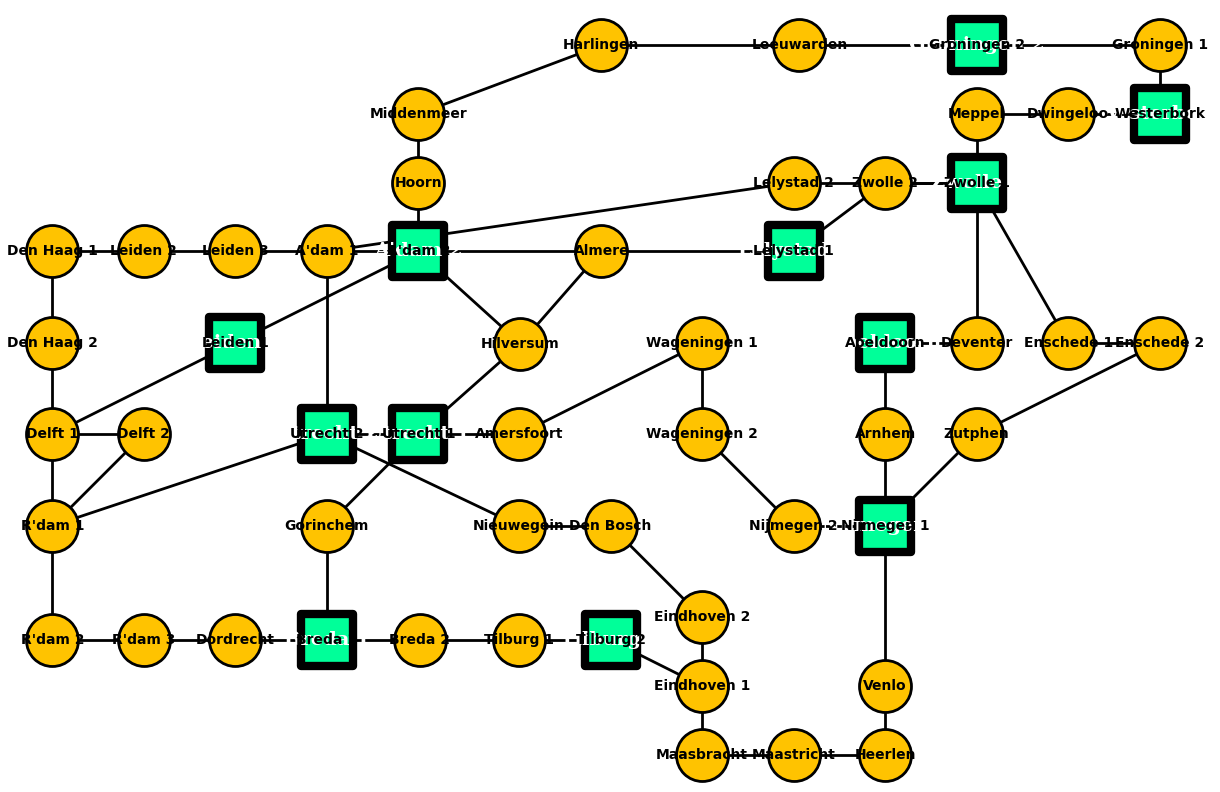}
\label{fig:lp60Surfnet}}

\subfigure[SCA-$\Lmax=130km$]{
\includegraphics[keepaspectratio=true,angle=0,width=.23\linewidth] {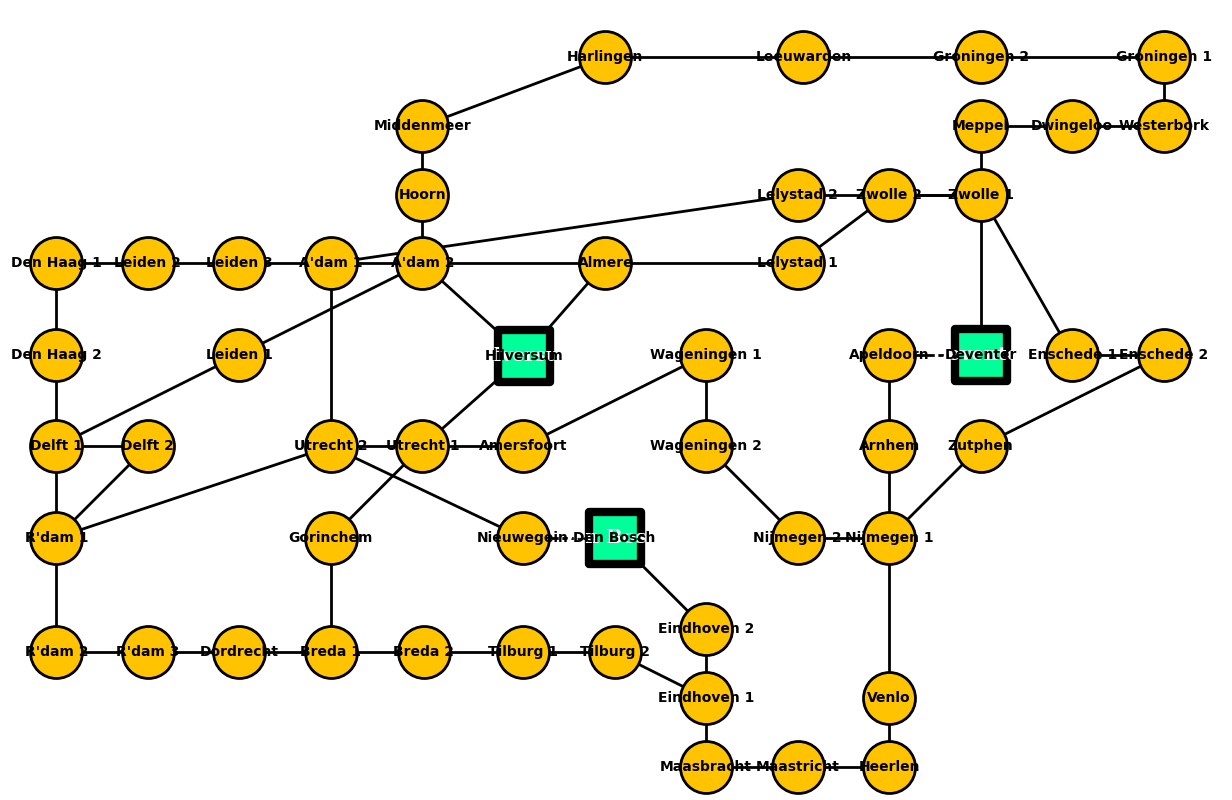}
\label{fig:greedy130Surfnet}}
\hspace{-4mm}
\subfigure[SCA-$\Lmax=100km$]{
\includegraphics[keepaspectratio=true,angle=0,width=.23\linewidth] {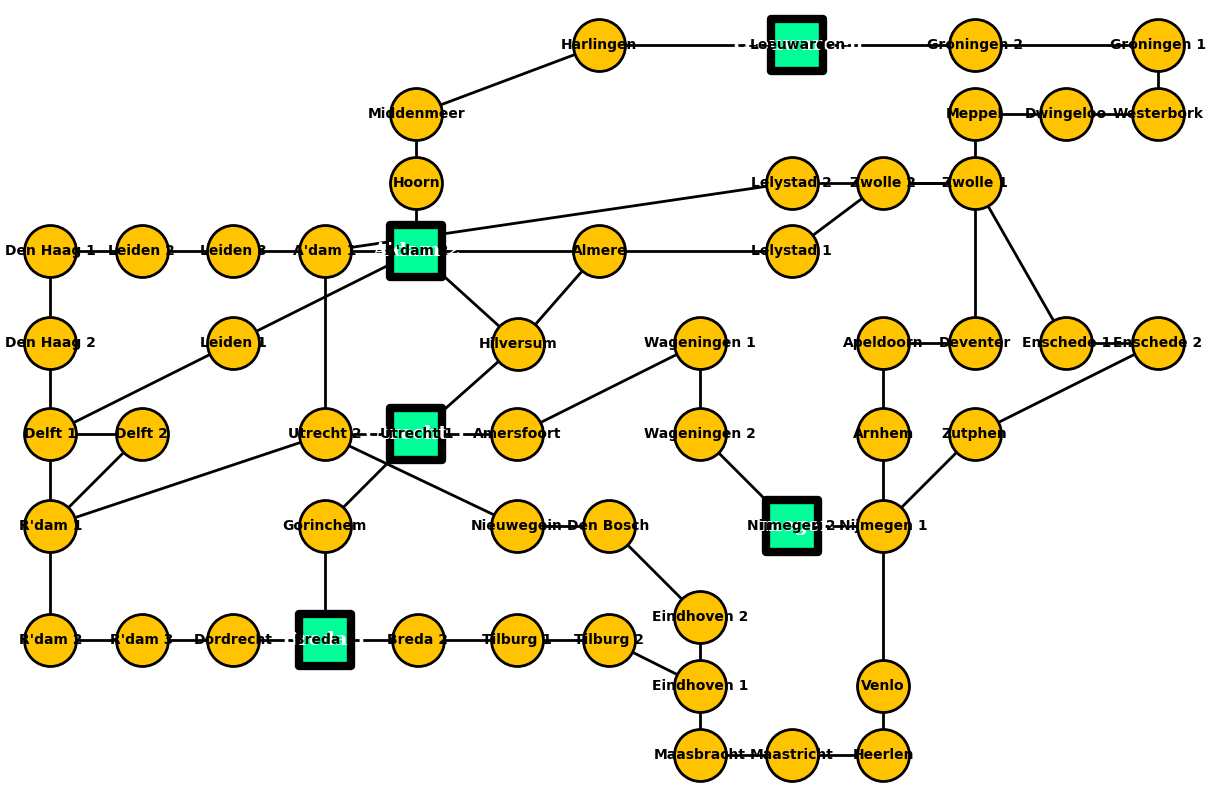}
\label{fig:greedy100Surfnet}}
\hspace{-4mm}
\subfigure[SCA-$\Lmax=80km$]{
\includegraphics[keepaspectratio=true,angle=0,width=.23\linewidth] {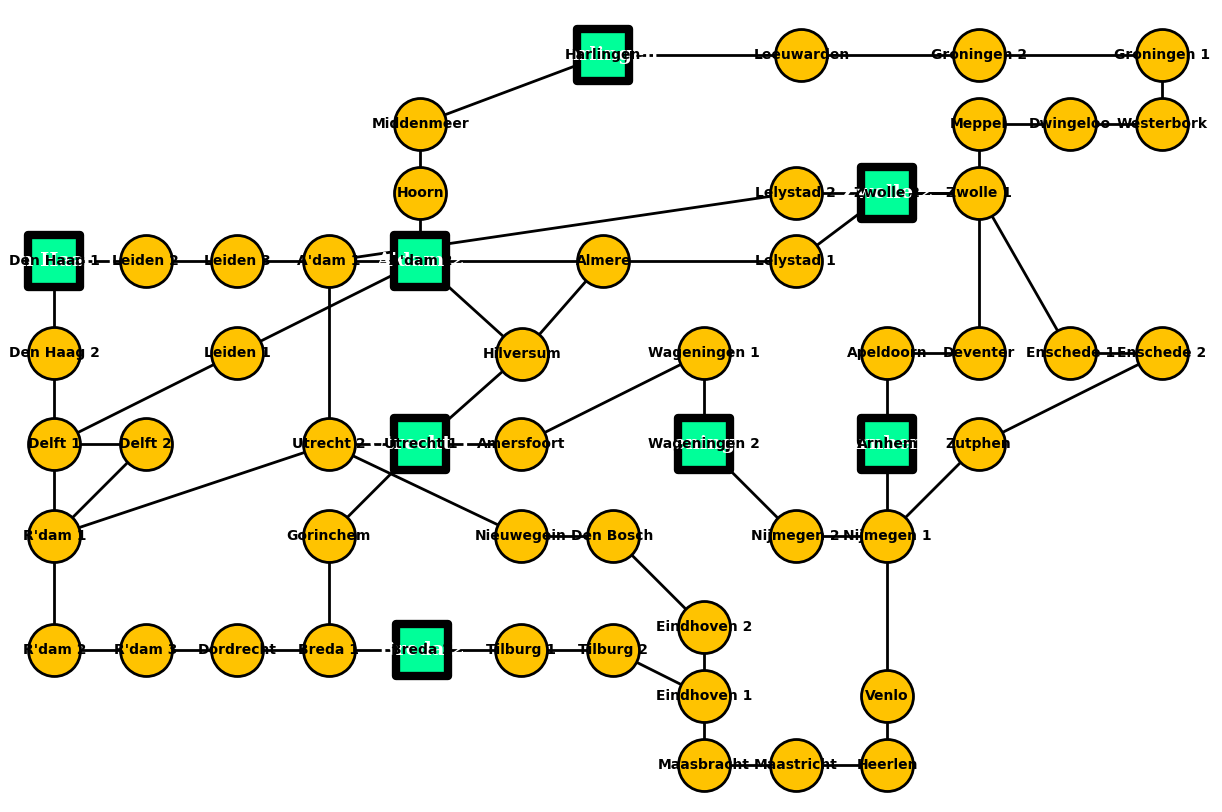}
\label{fig:greedy80Surfnet}}
\hspace{-4mm}
\subfigure[SCA-$\Lmax=60km$]{
\includegraphics[keepaspectratio=true,angle=0,width=.23\linewidth] {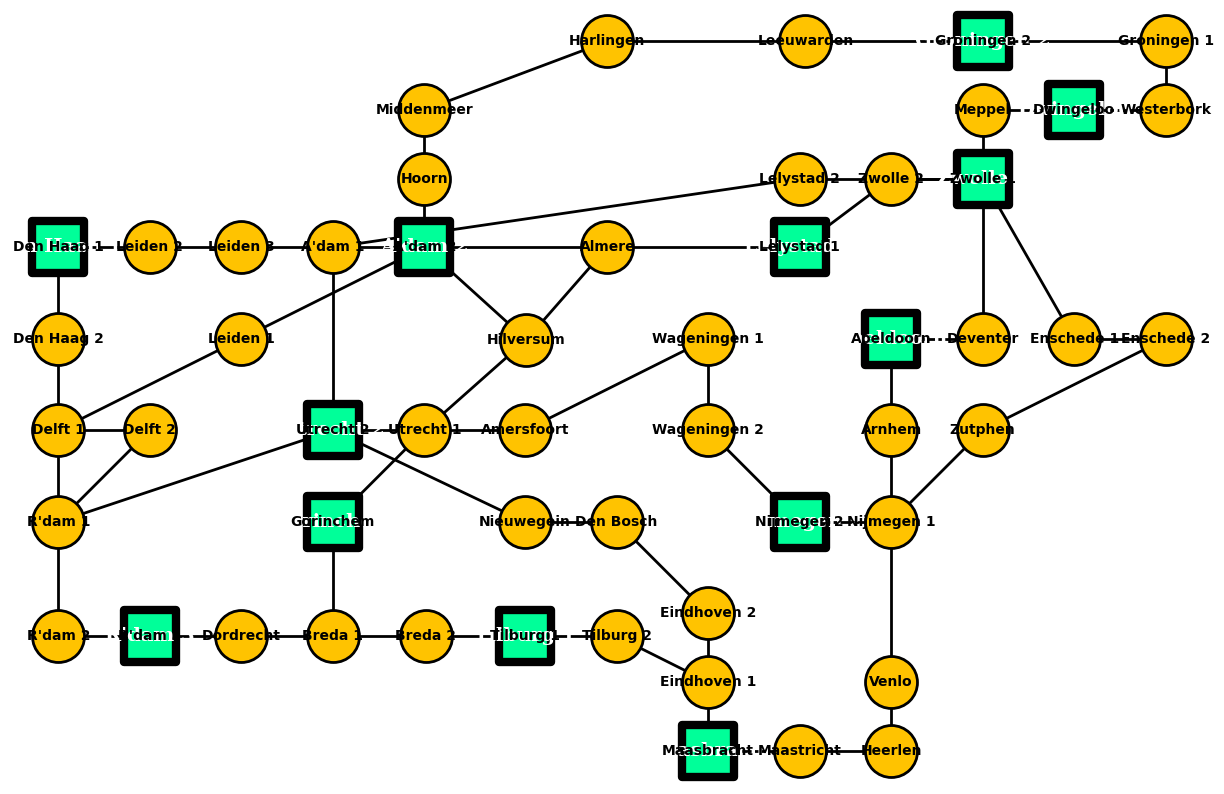}
\label{fig:greedy60Surfnet}}
\caption{The solutions for For SURFnet topology using Integer Linear Programming (ILP)~\cite{rabbie2022repallocLP} (a-d) and Single Center Approach (SCA).} 
\vspace{-6mm}
\label{fig:surfNetResult}
\end{center}
\end{figure*} 

Table \ref{tab:proposed_models_compare} displays the number of repeaters required by each method in SurfnetCore network for varying $\Lmax$ values when the robustness factor is set to $1$; i.e., no backup routes are selected for communication. Obviously, the ILP method results in the smallest number of quantum repeaters to create a feasible quantum communication path between all the hosts in the network. Among the heuristic methods we proposed, MCA-GP results in the largest number of quantum repeaters with as much as $2x$ of the ILP method. By relaxing the graph path selection requirement, MCA-Flex is able to reduce the number of quantum repeaters significantly compared to MCA-GP. This is mainly because of its ability to share the same intermediate nodes for the communication of multiple quantum repeaters as illustrated in Figure~\ref{fig:mca-flex-example}. Yet, it still requires almost twice as many quantum repeaters as the ILP method needs in some cases. As an example, the ILP returns $12$ nodes for quantum repeater deployment when $\Lmax=60$ km whereas MCA-Flex calculates $22$ quantum repeaters for the same condition.  

\begin{table}
\renewcommand{\arraystretch}{1.3}
\centering
\caption{Execution time comparison of SCA and ILP for SURFnet.}
\vspace{-2mm}
\label{tab:computation_time_comparison_surfnet}
\begingroup
\setlength{\tabcolsep}{2pt} 
\begin{tabular}{c c c }
\hline

Maximum transfer distance  &  \multicolumn{2}{c}{Execution Time (seconds)}\\
\cline{2-3}
  ($\Lmax$)  &  ILP & SCA \\
\hline
130 km & 13,481 & 0.28 \\
\hline
110 km& 30,300 & 0.31 \\
\hline
100 km& 78,944 & 0.36 \\

\hline
80 km& 167,769 & 0.39 \\
\hline
60 km& 167,496 & 0.39 \\
\hline
50 km& 218,085 & 0.40 \\
\hline
\end{tabular}
\endgroup
\vspace{-3mm}
\end{table}

\begin{table}
\renewcommand{\arraystretch}{1.3}
\centering
\caption{Comparison of quantum repeater count and execution time for Single Center Approach (SCA)  and Integer Linear Programming (ILP)~\cite{rabbie2022repallocLP} models for the random network.}
\vspace{-2mm}
\label{tab:computation_time_comparison_random_net}
\begingroup
\setlength{\tabcolsep}{2pt} 
\begin{tabular}{c c c c c}
\hline

Maximum transfer   &  \multicolumn{2}{c}{Quantum Repeaters} & \multicolumn{2}{c}{Execution Time (s)}\\
\cline{2-5}
  distance ($\Lmax$)  &  ILP & SCA &  ILP & SCA \\
\hline
0.9 &1 &1 & 0.98 & 0.015 \\
\hline
0.8 &2 &2 & 1.02 & 0.016 \\
\hline
0.7 &2 &2 & 0.86 & 0.018 \\
\hline
0.6 &3 &3 & 0.72 & 0.018 \\
\hline
0.5 &6 &7 & 0.55 & 0.017 \\

\hline
\end{tabular}
\endgroup
\end{table}

On the other hand, SCA yields the minimum number of repeaters among the heuristic methods. Even better, it returns at most one more quantum repeater than the ILP approach in all $\Lmax$ conditions. Figure \ref{fig:surfNetResult} compares the solutions found by SCA and ILP methods. We can observe that they choose the same nodes in many scenarios, showing the effectiveness of the SCA in comparison to ILP. Table~\ref{tab:computation_time_comparison_surfnet} shows the execution time of SCA and ILP methods under different $\Lmax$ values. It is clear that despite containing only $54$ nodes, the ILP method takes at least $13,481$ seconds (around $3.6$ hours) to find a solution. Its execution time reaches more than $60$ hours as the maximum qubit transfer distance is reduced since it needs to find more  nodes to deploy the quantum repeaters. On the other hand, the execution time of the heuristic model SCA remains less than one second in all $\Lmax$ values.

We also applied our approach to a random network topology used in~\cite{rabbie2022repallocLP}. The network consists of only well-connected $10$ nodes. The length of links is defined as one unit; thus, we set $\Lmax$ to multiple values between $0.5$ and $0.9$.  Table~\ref{tab:computation_time_comparison_random_net} shows the results for the estimated quantum repeater counts and execution times for SCA and ILP.  Both methods result in the same number of quantum repeaters except when $\Lmax=0.5$, for which SCA requires one more repeater than ILP.  Similar to SURFnet, the execution time of SCA is several orders lower than that of ILP.

\begin{table}
\renewcommand{\arraystretch}{1.3}
\centering
\caption{The performance of SCA for ESnet topology. Since the length of some links is larger than $\Lmax$, we augmented the network with new nodes for those links.}
\vspace{-2mm}
\label{tab:number_of_repeaters_ESnet}
\begingroup
\setlength{\tabcolsep}{2pt} 
\resizebox{.75\linewidth}{!}{
\begin{tabular}{c c c c c}
\hline
Maximum transfer   & \multicolumn{3}{c}{Quantum Repeaters} & Execution Time \\
\cline{2-4} 
  distance ($\Lmax$)  & New  & Existing & Total & (seconds)\\
\hline

300 & 38 & 21 & 59 & 1.98 \\
\hline
130 &  121 & 21 & 142 & 10.26 \\
\hline
110 &  153 & 21 & 174 & 14.49\\
\hline
100 &   168 & 21 & 189 & 20.67\\
\hline
80 &   220 & 20 & 242 & 33.48\\
\hline
60 &   300 & 23 & 323 & 62.15\\
\hline
50 &   336 & 22 & 358  & 90.91 \\
\hline
40 & 466 & 25 & 466 & 189.99\\
\hline
\end{tabular}}
\endgroup
\end{table}

In the case of ESnet network topology, the ILP approach could not find a solution by default since the length of some links is greater than $\Lmax$. To make it work, we augmented the network with new nodes for the links that are longer than $\Lmax$. When adding new nodes, we took the $\Lmax$ into consideration, so the number of new nodes is dependent on the value of $\Lmax$. As an example, we added $38$ new nodes with $\Lmax=300km$ and $121$ with $\Lmax=130$km. Hence, although the original topology only had $63$ nodes, the augmented version had more than $100$ nodes to accommodate the maximum qubit transfer distance. Despite running for several days, ILP did not find a solution, so we only executed SCA. Table~\ref{tab:number_of_repeaters_ESnet} displays the number of new repeaters required for different $\Lmax$ values. As expected, the number of new repeaters required increases as we increase the robustness parameter $K$. Yet, the execution time of SCA stayed in the order of a few minutes even when finding a solution for a network with more than $500$ vertices ($\Lmax$=40km), offering an extremely scalable solution for quantum repeater deployment modeling.

\textit{Failure Resistance:} Finally, we evaluated the performance of MCA-Flex, SCA, and ILP methods with a robustness requirement using the random network. Table~\ref{tab:random_network_robustness} shows the number of quantum repeaters estimated by each model with robustness factors of 1, 2, and 3. Robustness factor 1 refers to the scenario with no backup routes, whereas $2$ and $3$ refer to the existence of one and two backup routes between every possible pair of node communication. As expected, increasing the number of backup routes (i.e., the value of $K$) requires more nodes to be used for quantum repeater deployment. As an example, while only $2$ nodes are selected when $\Lmax=0.7$ with $K=1$, $7-8$ quantum repeaters are needed when $K=3$. On the other hand, similar to previous results, SCA requires at most one more quantum repeater than ILP in all cases whereas MCA-Flex requires a significantly higher number of  quantum repeaters to provide a similar robustness functionality.
\begin{table}[t]
\renewcommand{\arraystretch}{1.1}
\centering
\caption{Quantum repeater estimations considering link/node failures. Up to $K-1$ failures can be resisted.}
\vspace{-2mm}
\label{tab:random_network_robustness}
\begingroup
\setlength{\tabcolsep}{1pt} 
\resizebox{.75\linewidth}{!}{
\begin{tabular}{c c c c c}
\hline
Maximum transfer & Total  &  \multicolumn{3}{c}{Quantum Repeater Count}\\
\cline{3-5}
$\Lmax$ (km) & Routes (K) & MCA-Flex  & SCA  & ILP\\
\hline
\multirow{3}{*}{{0.9}} & 1 & 1  & 1 & 1\\
\cline{2-5} & 2 & 2  & 2 & 2\\
\cline{2-5} & 3 & 3  & 3 & 3\\
\hline
\multirow{3}{*}{{0.8}} & 1 & 3  & 2 & 2\\
\cline{2-5} & 2 & 7  & 4 & 4\\
\cline{2-5} & 3 & 9  & 6 & 6\\
\hline
\multirow{3}{*}{{0.7}} & 1 & 4  & 2 & 2\\
\cline{2-5} & 2 & 9  & 4 & 4\\
\cline{2-5} & 3 & 12  & 8 & 7\\
\hline
\multirow{2}{*}{{0.6}} & 1 & 7  & 3 & 3\\
\cline{2-5} & 2 & 11  & 8 & 7\\
\hline
\end{tabular}}
\endgroup
\end{table}

%% file: conclusion.tex
Quantum repeaters are essential components of the quantum internet as qubits cannot be transferred to long distances without losing their coherence. In this paper, we propose two heuristic algorithms for quantum repeater deployment modeling and show that they attain near-optimal solutions compared to the linear programming-based solution while reducing the execution times from days to seconds. 

In future work, we plan to consider the capacity of quantum repeaters to find solutions that can allow all hosts to communicate at the same time. To address this limitation, we can calculate the maximum capacity required by the number of hosts that will pass through each quantum repeater to communicate with other hosts. Then, we can extend the robustness-based solution to allow multiple routes to be created between endpoints.

%% file: main.bbl
\begin{thebibliography}{10}
\providecommand{\url}[1]{#1}
\csname url@samestyle\endcsname
\providecommand{\newblock}{\relax}
\providecommand{\bibinfo}[2]{#2}
\providecommand{\BIBentrySTDinterwordspacing}{\spaceskip=0pt\relax}
\providecommand{\BIBentryALTinterwordstretchfactor}{4}
\providecommand{\BIBentryALTinterwordspacing}{\spaceskip=\fontdimen2\font plus
\BIBentryALTinterwordstretchfactor\fontdimen3\font minus
  \fontdimen4\font\relax}
\providecommand{\BIBforeignlanguage}[2]{{%
\expandafter\ifx\csname l@#1\endcsname\relax
\typeout{** WARNING: IEEEtran.bst: No hyphenation pattern has been}%
\typeout{** loaded for the language `#1'. Using the pattern for}%
\typeout{** the default language instead.}%
\else
\language=\csname l@#1\endcsname
\fi
#2}}
\providecommand{\BIBdecl}{\relax}
\BIBdecl

\bibitem{ekert1991quantumkd}
A.~K. Ekert, ``Quantum cryptography based on bell’s theorem,'' \emph{Physical
  review letters}, vol.~67, no.~6, p. 661, 1991.

\bibitem{bennett2020quantumkd}
C.~H. Bennett and G.~Brassard, ``Quantum cryptography: Public key distribution
  and coin tossing,'' \emph{arXiv preprint arXiv:2003.06557}, 2020.

\bibitem{pirandola2020advancescrypto}
S.~Pirandola, U.~L. Andersen, L.~Banchi, M.~Berta, D.~Bunandar, R.~Colbeck,
  D.~Englund, T.~Gehring, C.~Lupo, C.~Ottaviani \emph{et~al.}, ``Advances in
  quantum cryptography,'' \emph{Advances in optics and photonics}, vol.~12,
  no.~4, pp. 1012--1236, 2020.

\bibitem{jozsa2000quantumclock}
R.~Jozsa, D.~S. Abrams, J.~P. Dowling, and C.~P. Williams, ``Quantum clock
  synchronization based on shared prior entanglement,'' \emph{Physical Review
  Letters}, vol.~85, no.~9, p. 2010, 2000.

\bibitem{chuang2000quantumclock}
I.~L. Chuang, ``Quantum algorithm for distributed clock synchronization,''
  \emph{Physical review letters}, vol.~85, no.~9, p. 2006, 2000.

\bibitem{giovannetti2001quantumclock}
V.~Giovannetti, S.~Lloyd, and L.~Maccone, ``Quantum-enhanced positioning and
  clock synchronization,'' \emph{Nature}, vol. 412, no. 6845, pp. 417--419,
  2001.

\bibitem{buhrman2003distributed}
H.~Buhrman and H.~R{\"o}hrig, ``Distributed quantum computing,'' in
  \emph{Mathematical Foundations of Computer Science 2003: 28th International
  Symposium, MFCS 2003, Bratislava, Slovakia, August 25-29, 2003. Proceedings
  28}.\hskip 1em plus 0.5em minus 0.4em\relax Springer, 2003, pp. 1--20.

\bibitem{beals2013efficientdistributed}
R.~Beals, S.~Brierley, O.~Gray, A.~W. Harrow, S.~Kutin, N.~Linden, D.~Shepherd,
  and M.~Stather, ``Efficient distributed quantum computing,''
  \emph{Proceedings of the Royal Society A: Mathematical, Physical and
  Engineering Sciences}, vol. 469, no. 2153, p. 20120686, 2013.

\bibitem{cacciapuoti2019quantumdistributed}
A.~S. Cacciapuoti, M.~Caleffi, F.~Tafuri, F.~S. Cataliotti, S.~Gherardini, and
  G.~Bianchi, ``Quantum internet: networking challenges in distributed quantum
  computing,'' \emph{IEEE Network}, vol.~34, no.~1, pp. 137--143, 2019.

\bibitem{bergli2009decoherence}
J.~Bergli, Y.~M. Galperin, and B.~Altshuler, ``Decoherence in qubits due to
  low-frequency noise,'' \emph{New Journal of Physics}, vol.~11, no.~2, p.
  025002, 2009.

\bibitem{saki2019studydecoherence}
A.~A. Saki, M.~Alam, and S.~Ghosh, ``Study of decoherence in quantum computers:
  A circuit-design perspective,'' \emph{arXiv preprint arXiv:1904.04323}, 2019.

\bibitem{helm2009quantumdecoherence}
J.~Helm and W.~T. Strunz, ``Quantum decoherence of two qubits,'' \emph{Physical
  Review A}, vol.~80, no.~4, p. 042108, 2009.

\bibitem{rabbie2022repallocLP}
J.~Rabbie, K.~Chakraborty, G.~Avis, and S.~Wehner, ``Designing quantum networks
  using preexisting infrastructure,'' \emph{npj Quantum Information}, vol.~8,
  no.~1, p.~5, 2022.

\bibitem{chakraborty2019distributed}
K.~Chakraborty, F.~Rozpedek, A.~Dahlberg, and S.~Wehner, ``Distributed routing
  in a quantum internet,'' \emph{arXiv preprint arXiv:1907.11630}, 2019.

\bibitem{yao2021optimaldeploymentdesign}
J.~Yao, K.~Zou, D.~Li, and Z.~Jiang, ``Optimal deployment design of repeaters
  and memories in quantum networks,'' in \emph{2021 IEEE 23rd Int Conf on High
  Performance Computing \& Communications; 7th Int Conf on Data Science \&
  Systems; 19th Int Conf on Smart City; 7th Int Conf on Dependability in
  Sensor, Cloud \& Big Data Systems \& Application
  (HPCC/DSS/SmartCity/DependSys)}.\hskip 1em plus 0.5em minus 0.4em\relax IEEE,
  2021, pp. 361--368.

\bibitem{liorni2021quantumrepinspace}
C.~Liorni, H.~Kampermann, and D.~Bru{\ss}, ``Quantum repeaters in space,''
  \emph{New Journal of Physics}, vol.~23, no.~5, p. 053021, 2021.

\bibitem{dahlberg2019linklayerprot}
A.~Dahlberg, M.~Skrzypczyk, T.~Coopmans, L.~Wubben, F.~Rozp{\k{e}}dek,
  M.~Pompili, A.~Stolk, P.~Pawe{\l}czak, R.~Knegjens, J.~de~Oliveira~Filho
  \emph{et~al.}, ``A link layer protocol for quantum networks,'' in
  \emph{Proceedings of the ACM special interest group on data communication},
  2019, pp. 159--173.

\bibitem{kozlowski2020designingqnprot}
W.~Kozlowski, A.~Dahlberg, and S.~Wehner, ``Designing a quantum network
  protocol,'' in \emph{Proceedings of the 16th international conference on
  emerging networking experiments and technologies}, 2020, pp. 1--16.

\bibitem{yang2022online}
L.~Yang, Y.~Zhao, H.~Xu, and C.~Qiao, ``Online entanglement routing in quantum
  networks,'' in \emph{2022 IEEE/ACM 30th International Symposium on Quality of
  Service (IWQoS)}.\hskip 1em plus 0.5em minus 0.4em\relax IEEE, 2022, pp.
  1--10.

\bibitem{zhao2021redundant}
Y.~Zhao and C.~Qiao, ``Redundant entanglement provisioning and selection for
  throughput maximization in quantum networks,'' in \emph{IEEE INFOCOM
  2021-IEEE Conference on Computer Communications}.\hskip 1em plus 0.5em minus
  0.4em\relax IEEE, 2021, pp. 1--10.

\bibitem{zhao2022segmented}
G.~Zhao, J.~Wang, Y.~Zhao, H.~Xu, and C.~Qiao, ``Segmented entanglement
  establishment for throughput maximization in quantum networks,'' in
  \emph{2022 IEEE 42nd International Conference on Distributed Computing
  Systems (ICDCS)}.\hskip 1em plus 0.5em minus 0.4em\relax IEEE, 2022, pp.
  45--55.

\bibitem{lam2014electriccomplexity}
A.~Y. Lam, Y.-W. Leung, and X.~Chu, ``Electric vehicle charging station
  placement: Formulation, complexity, and solutions,'' \emph{IEEE Transactions
  on Smart Grid}, vol.~5, no.~6, pp. 2846--2856, 2014.

\bibitem{kruskal1956shortest}
J.~B. Kruskal, ``On the shortest spanning subtree of a graph and the traveling
  salesman problem,'' \emph{Proceedings of the American Mathematical society},
  vol.~7, no.~1, pp. 48--50, 1956.

\bibitem{surfnet}
``{SURF},'' 2023, https://www.surf.nl/en.

\bibitem{esnet}
``{Energy Sciences Network},'' 2023, https://www.es.net/.

\end{thebibliography}
